\documentclass[12pt]{article}
\usepackage[margin=1in]{geometry}
\usepackage[square,numbers]{natbib}
\usepackage{amsmath,amsfonts,amsthm,graphicx}
\usepackage{algorithm,algpseudocode}

\begin{document}

\title{Nonparametric Empirical Bayes Biomarker Imputation and Estimation}
\author{Alton Barbehenn and Sihai Dave Zhao}
\maketitle

\begin{abstract}
Biomarkers are often measured in bulk to diagnose patients, monitor patient conditions, and research novel drug pathways. The measurement of these biomarkers often suffers from detection limits that result in missing and untrustworthy measurements. Frequently, missing biomarkers are imputed so that down-stream analysis can be conducted with modern statistical methods that cannot normally handle data subject to informative censoring. This work develops an empirical Bayes $g$-modeling method for imputing and denoising biomarker measurements. We establish superior estimation properties compared to popular methods in simulations and demonstrate the utility of the estimated biomarker measurements for down-stream analysis.

\end{abstract}

\section{Introduction}
\label{sec:intro}

The measurement of biomarkers is a fundamental task in many modern clinical and biomedical studies. Biomarkers are measurable indicators of biological or pathological processes that can be used to provide important insights into disease diagnosis, monitoring, and treatment. However, the measurement of biomarkers is not without challenges. Many medical studies often have small sample sizes until a phenomena is well understood so efficient data use is essential \cite{GenIMS}. Beyond the usual measurement errors, limitations in laboratory collection and measurement procedures can result in detection limits for biomarker measurement. Detection limits often manifest as left-censoring, right-censoring, and in cases such as rounding, interval-censoring. For example, in left-censoring may occur when it is impossible to determine if a biomarker is present in small concentrations or simply absent. Detection limits produce missing not at random data, such forms of missingness are non-ignorable and failing to properly handle the missingness can introduce bias in statistical procedures \cite{Troyanskaya_2001, TaylorLeiserowitzKim2013, KarpievitchDabneySmith2012}. Properly accounting for detection limits is important in many applications, such as the measurement of IL-6 and IL-10 cytokines for sepsis \cite{GenIMS, LeeKongWeissfeld2012} or CD4$^+$ T-lymphocytes for human immunodeficiency virus \cite{Mellors_1996, Lynn2001}.

Much of the work handling biomarker measurements suffering from detection limits can be classified as either directly estimating the missing biomarker or modifying the analysis to account for missing biomarker measurements. These approaches can often be thought of as regression problems that either treat the measured biomarker as a censored response or predictor, respectively \cite{LeeKongWeissfeld2012}. In this work, we focus on directly estimating the missing biomarkers so that complex downstream analysis can be easily conducted using modern machine learning and data mining methods. Because the data are missing not at random, the usual approach of only utilizing observed data is not viable \cite{RubinLittle2020}, instead we require an explicit model of the missing mechanism.

Popular methods for estimating missing biomarker measurements span a wide range of complexities. The most basic approach to handle low detection limits are the so-called ``fill-in'' methods. These methods estimate the missing measurement with some constant function of the detection limit based on the distribution of the censored tail \cite{LeeKongWeissfeld2012}. For example, if a nonnegative concentration falls below a limit-of-detection (LOD), it may be estimated as $LOD$, $LOD/2$, or $LOD/\sqrt{2}$. These methods are easy to implement but ignore the relationship between biomarkers and lack variability that may be crucial for latter analysis. Regression based approaches offer a natural extension to the fill-in methods; rather than relying on the censoring mechanism alone, these methods use every measurement of a biomarker to model the distribution of values \cite{Lynn2001}. Covariates, either demographic or fully observed biomarkers, can be included in the regression model to account for additional variability in the data \cite{Lubin_2004}. Once the regression model is fit, samples can be conditionally drawn to recreate the full data variability \cite{Lynn2001, Lubin_2004, Wei_GSimp2018}. Nearest neighbor methods offer a nonparametric regression alternative for estimating the missing biomarkers \cite{Shah_2017}. Once the measurements are standardized, the nearest neighbors can be computed as nearest biomarkers or nearest patients. Nearest patients is generally preferable for biomarker estimation because it can capture complex relationships between many biomarkers. Unfortunately, by construction, nearest neighbor methods cannot impute biomarkers whose values lie outside the observed range. Many other nonparametric methods such as random forests \cite{StekhovenBuhlmann2011} and singular value decomposition \cite{Hastie_1999} have been proposed but they often struggle in the missing not at random setting that we are studying \cite{Lazar_2016, Wei_2018}. Many of these methods modify a likelihood to handle the informative censoring. If necessary, a modified Box-Cox transformation can be employed to ensure that the data have nearly a Gaussian distribution subject to any detection limits before using any imputation method the builds on the Gaussian model \cite{HanKronmal2004}.

We are motivated by applications where many biomarkers are measured simultaneously so that their combination can be used to diagnose and monitor one or more conditions \cite{French_2016, Zhou_2021}. These data are often acquired with tools such as mass spectroscopy \cite{TaylorLeiserowitzKim2013} or flow cytometry \cite{Nolan2022}. In these cases, the relationships between biomarkers can be leveraged to estimate missing measurements \cite{Shah_2017}; however, the introduction of additional censored biomarkers increases the difficulty of the estimation problem.

In this paper we propose addressing these difficulties by developing a nonparametric empirical Bayes method. Empirical Bayes methods estimate the Bayes optimal regression function for denoising biomakers and, in doing so, provide a very powerful tool for simultaneous estimation problems \cite{Fourdrinier2018, JiangZhang2009, JamesStein1961}. The empirical Bayes approach assumes that the true biomaker values are drawn independently from some unknown prior, $g$, and the corresponding observations are drawn from a known likelihood \cite{Robbins1956}. Under this Bayesian model, the posterior mean is usual used as the estimate for each biomaker and parameters required to compute the posterior mean are estimated from the observed marginal distribution \cite{JiangZhang2009, Efron2019}. There are, of course, many other ways to regularize models to improve estimation such as ridge and LASSO penalties \cite{Hastie_2009}; however, we prefer empirical Bayes methods because they are tuning-parameter free \cite{KoenkerMizera2014}, easy to implement, and have strong theoretical guarantees \cite{JiangZhang2009, JamesStein1961, Robbins1951, SoloffGuntuboyinaSen2021}. We note that empirical Bayes can be seen as a self-supervised regression problem \cite{BarbehennZhao2023}, as such, it bridges the conceptual gap between treating the biomarker as a response and a predictor in the regression problems.

In this work we follow the nonparametric empirical Bayes $g$-modeling framework \cite{KoenkerMizera2014, Efron2014}. This approach assumes no structure on the prior, $g$, and produces an estimated prior, $\hat{g}$, using nonparametric maximum marginal likelihood estimation \cite{KieferWolfowitz1956}. The posterior mean is estimated using $\hat{g}$ and the known likelihood. In cases where there is no corresponding biomarker measurement, for example when there is censoring due to a detection limit, we can still compute the posterior mean given that the biomarker measurement fell within a specific range.

Our key insight is that because popular biomaker estimation methods have established likelihoods for censored biomaker measurements \cite{Lynn2001, HanKronmal2004, LylesFanChuachoowong2001}, nonparametric empirical Bayes methods can be directly employed to improve the simultaneous estimation of biomakers without requiring additional domain knowledge or tuning. Using nonparametric empirical Bayes $g$-modeling formulation, we show superior estimation and imputation performance in simulations based on real data compared when compared to popular methods. We provide an open-source R package \texttt{ebTobit} (https://github.com/barbehenna/ebTobit) for implementing our proposed methods.

\section{Empirical Bayes Matrix Imputation}
\label{sec:method}

\subsection{Methodology}
\label{sec1:methodology}

We are interested in estimating and imputing $p$ biomarkers from each of $n$ patients. Here, true biomarker values of patient $i$ are denoted as independent samples $(\theta_{i1}, \dots \theta_{ip}) \sim g$ on $\mathbb{R}^p$. Assume we observe intervals $[L_{ij}, R_{ij}]$ for each patient $i = 1, \dots, n$ and biomarker $j = 1, \dots, p$. When $L_{ij} = R_{ij}$, a noisy observation of the $\theta_{ij}$ is directly measured; we assume that the error is normally distributed so that the contribution to the likelihood is $\phi_{\sigma_{ij}}(L_{ij} - \theta_{ij})$, where $\phi_\sigma(\cdot)$ denotes the Gaussian density function with variance $\sigma^2 > 0$. When $L_{ij} < R_{ij}$, the observation is interval censored and the contribution to the likelihood is $\Phi_{\sigma_{ij}}(R_{ij} - \theta_{ij}) - \Phi_{\sigma_{ij}}(L_{ij} - \theta_{ij})$, where $\Phi_\sigma(\cdot)$ denotes the Gaussian distribution function with variance $\sigma^2 > 0$. For example, if a biomarker's concentration falls below a lower limit-of-detection ($LOD$), a direct measurement is not possible; however, because concentrations are non-negative, we observe the interval $[0, LOD]$. If a biomarker is successfully measured, $[L_{ij}, R_{ij}]$ contains a single noisy point estimate of $\theta_{ij}$. This data structure is sometimes referred to as general partly interval-censored data \cite{Huang1999}. For most of our methodology we focus on the case where $\sigma_{ij}^2$ are known; however, methods allowing for the joint estimation of $\theta_{ij}$ and $\sigma_{ij}^2$ are discussed below. We represent the full set of observations, in matrix form, as:
\begin{align*}
	\mathbf{L} = \begin{bmatrix}
	L_{11} & \dots  & L_{1p} \\
	L_{21} & \dots  & L_{2p} \\
	\vdots & \ddots & \vdots \\
	L_{n1} & \dots  & L_{np} \\
	\end{bmatrix}
	\qquad \text{and} \qquad
	\mathbf{R} = \begin{bmatrix}
	R_{11} & \dots  & R_{1p} \\
	R_{21} & \dots  & R_{2p} \\
	\vdots & \ddots & \vdots \\
	R_{n1} & \dots  & R_{np} \\
	\end{bmatrix} .
\end{align*}
We use the notation $L_{i \cdot}$ and $L_{\cdot j}$ to denote the row vector $(L_{i1}, \dots, L_{ip})$ and the column vector $(L_{1j}, \dots, L_{nj})$, respectively.

Under our Bayesian model, a natural estimator of $\theta_{i \cdot}$ is the posterior mean $E( \theta_{i\cdot} \mid L_{i\cdot}, R_{i\cdot})$. Observe that the posterior mean is given by
\begin{equation}
	\label{eq:posteriormean}
	E( \theta_{i\cdot} \mid L_{i\cdot}, R_{i\cdot})
		= \frac{ \int_{\mathbb{R}^p} t P(L_{i\cdot}, R_{i\cdot} \mid \theta_{i\cdot} = t) ~ d g(t) }{ \int_{\mathbb{R}^p} P(L_{i\cdot}, R_{i\cdot} \mid \theta_{i\cdot} = t) ~ d g(t) } ,
\end{equation}
where the likelihood $P(L_{i \cdot}, R_{i \cdot} \mid \theta_{i \cdot})$ is given by
\begin{align}
	\label{eq:tobitlik}
	P(L_{i\cdot}, R_{i\cdot} \mid \theta_{i\cdot}) 
		&= \prod_{j=1}^p P(L_{ij}, R_{ij} \mid \theta_{ij}) \nonumber \\
		&= \prod_{j=1}^p \left\{ \phi_{\sigma_{ij}}(L_{ij} - \theta_{ij}) \right\}^{1(L_{ij} = R_{ij})} \left\{ \Phi_{\sigma_{ij}}(R_{ij} - \theta_{ij}) - \Phi_{\sigma_{ij}}(L_{ij} - \theta_{ij}) \right\}^{1(L_{ij} < R_{ij})} .
\end{align}
Each term in the product \eqref{eq:tobitlik} is a Tobit likelihood with $\sigma_{ij}^2$ variance \cite{Lubin_2004, Tobin1958, PerssonRootzen1977, Amemiya1973}. We note that underlying physiological conditions may manifest as dependent biomarker expressions; accordingly, we will not impose any independence structures on the prior, $g$, such as a mean field approximation. Empirical Bayes $g$-modeling suggests that estimating $g$ from the data and plugging $\hat{g}$ into \eqref{eq:posteriormean} results in a good estimator \cite{Efron2014}. 

When there are at least two measurements for every $\theta_{ij}$, empirical Bayes modeling can be extended to estimate both means and variances \cite{GuKoenker2017}. Additional measurements of $\theta_{ij}$ are often called technical replicates; including replicates adds extra overhead to the measurement process but, by allowing for the estimation of the noise levels, we make the results more robust to misspecified noise models. The simplest empirical Bayes approach is to assume a prior on the means and variances of each patient's biomarker measurements, $g(\theta_1, \dots, \theta_p, \sigma^2_1, \dots, \sigma^2_p)$, then specify the appropriate likelihood and proceed as we have previously in this section. The increased dimensionality of the prior can make estimation more difficult \cite{GuKoenker2017}. Many simplifying assumptions can be made on the distribution to accommodate different physical models. For example, we could continue to assume that the biomarker mean values are arbitrarily related but also assume that the variance of each measurement only depends on the value of the biomarker being measured. This model results in the following Bayesian decomposition of the prior: 
\begin{align*}
	g(\theta_1, \dots, \theta_p, \sigma^2_1, \dots, \sigma^2_p)
		&= g(\theta_1, \dots, \theta_p) g(\sigma^2_1, \dots, \sigma^2_p \mid \theta_1, \dots, \theta_p) \\
		&= g(\theta_1, \dots, \theta_p) \prod_{j=1}^p g(\sigma^2_j \mid \theta_j) .
\end{align*}
In this Bayesian decomposition, we reduce the prior's complexity by arguing for conditional independence of the variances. We note that each of the $g(\sigma^2_j \mid \theta_j)$ can be learned as a regression problem in independent control assays or specified to match a physical model. We stress that modeling both location and scale parameters is not possible without measurement replicates and that the choice of model should reflect the needs of the specific assays used.

\subsection{Implementation}
\label{sec1:implementation}

Estimating the prior, $g$, can be done in many ways. Proceeding with standard nonparametric empirical Bayes $g$-modeling arguments, we model $g$ in the space of all distributions on $\mathbb{R}^p$ and estimate it using maximum marginal likelihood:
\begin{align}
	\label{eq:maxmarglik}
\hat{g} 
	&= \arg\max_{g} \sum_{i=1}^n \log P(L_{i\cdot}, R_{i\cdot}) \nonumber \\
	&= \arg\max_{g} \sum_{i=1}^n \log \int_{\mathbb{R}^p} P(L_{i\cdot}, R_{i\cdot} \mid \theta_{i\cdot} = t) ~ d g(t) .
\end{align}
This optimization problem is concave, but infinite-dimensional. Fortunately, Carath\'eodory’s theorem of convex hulls \cite{JiangZhang2009, Caratheodory1911} ensures that there is a discrete distribution, $g^{*}$, with at most $n+1$ support points that solves \eqref{eq:maxmarglik}. Accordingly, we simplify the infinite-dimensional optimization problem, \eqref{eq:maxmarglik}, by focusing on distributions supported on a finite set of $m > 0$ support points $t_1, \dots, t_m \in \mathbb{R}^p$. After fixing the $m$ support points, $g$ has the form $g(t) = \sum_{k=1}^m w_k \delta_{t_k}(t)$, where each $w_k \geq 0$ and $\sum_{k=1}^m w_k = 1$. The optimization problem is then \cite{JiangZhang2009, KoenkerMizera2014}: 
\begin{equation}
	\label{eq:NPMLE}
	\hat{g} = \arg\max_{\mathbf{w} : w_k \geq 0, \sum_{k=1}^m w_k = 1} \sum_{i=1}^n \log \sum_{k=1}^m w_k P(L_{i\cdot}, R_{i\cdot} \mid \theta_{i\cdot} = t_k)
\end{equation}
With fixed support points, only $w_1, \dots, w_m$ need to be estimated, this means that \eqref{eq:NPMLE} is a finite-dimensional, convex optimization problem that can be solved by many optimization libraries \cite{KoenkerMizera2014}. It is possible to simultaneously estimate both $t_k$ and $w_k$; however, the resulting optimization problem is non-convex.

Selecting the support points for multi-dimensional $g$ is a nontrivial task for which there is no good solution. The optimal support points for the empirical Bayes problem are known to be $\theta_{i\cdot}$ themselves \cite{JiangZhang2009}; however, since the $\theta_{i\cdot}$ are unknown in practice, another method must be employed to specify support points with minimal misspecification error. Most approaches to this problem either use a regular grid over the range of the observations \cite{JiangZhang2009, KoenkerMizera2014, SoloffGuntuboyinaSen2021} or the observations themselves \cite{SahaGuntuboyina2020} as support points for $g$. The later method is often referred to as the ``exemplar method''.

Standard methods for support point selection do not perform well for our problem. The regular grid method suffers from the curse-of-dimensionally: as $p$ increases, exponentially more support points are required to ensure closeness to the optimal support points. In practice, a dense grid with hundreds of support points per axis is not computationally feasible if $p$ is greater than 3 or 4. The exemplar method offers direct relief to the curse-of-dimensionality by using the observations as support points, thus avoiding the dependence of dimension on the support size. Unfortunately, in our application, we do not have direct measurements of every $\theta_{i \cdot}$ because of censoring, so we cannot directly apply the exemplar method.

Briefly, we note that the exemplar method can be generalized to handle our censored observations by using the maximum likelihood of each $\theta_{i\cdot}$ as support points. Under the Tobit likelihood \eqref{eq:tobitlik}, when $L_{ij}$ and $R_{ij}$ are finite, the maximum likelihood estimate of $\theta_{ij}$ is:
\begin{equation}
	\label{eq:mle}
	\hat{\theta}_{ij} = \frac{L_{ij} + R_{ij}}{2} .
\end{equation}
Using $\hat{\theta}_{i \cdot}$ as generalized exemplar support does not perform well in our simulations, see Appendix \ref{app:sims}. We note that when $R_{ij} - L_{ij}$ is large compared to $\sigma_{ij}$, the corresponding support point $\hat{\theta}_{i \cdot}$ may be far from the optimal support point $\theta_{i \cdot}$. For example, if $L_{ij} = 0$, $R_{ij} = 1000$, and $\sigma_{ij}^2 = 1$, then, on average, $\hat{\theta}_{ij}=500$ is a much worse estimate of $\theta_{ij}$ than a sample from $N(\theta_{ij}, \sigma_{ij}^2 = 1)$ for most $\theta_{ij}$. Additionally, using \eqref{eq:mle} reduces to the usual exemplar support when there is no censoring. We finally note that when there is a common censoring interval, \ref{eq:mle} is an example of a fill-in method \cite{LeeKongWeissfeld2012}.

The key insight of the exemplar method is that samples from the uncensored marginal distribution are likely to be close to the oracle support points \cite{SahaGuntuboyina2020}. This idea inspires us to develop support point selection methods that draw on sampling algorithms; samples from the uncensored marginal distribution are likely to be good support points. Sampling algorithms are not new to biomarker imputation; both Gibbs sampling \cite{LeeKongWeissfeld2012} and bootstrap sampling \cite{Lubin_2004} schemes have been used to impute missing values given fully observed covariates under the Tobit regression model.


We construct a novel, heuristic algorithm, for empirical Bayes matrix estimation under a Tobit likelihood, called ``EBM-Tobit''. Our key insight is that if we know the prior, $g$, then sampling from the uncensored marginal distribution according to our Bayesian model is easy. Additionally, the exemplar method suggests that we only need the number of samples from the uncensored marginal to grow like $n$, thus avoiding the curse-of-dimensionality. Algorithm \ref{alg} illustrates our proposed fitting scheme that alternates between estimating $g$ and using sampling support points from an approximate, uncensored marginal distribution. Many methods can be used to produce a final estimate, for example, one could simply use the final estimated prior along with \eqref{eq:posteriormean}. In Algorithm \ref{alg}, we draw inspiration from standard sampling methods and average multiple estimated posterior means to be used as the final estimate.

\begin{algorithm}
\caption{An algorithm to perform support point selection and compute ``EBM-Tobit''.}
\label{alg}
\begin{algorithmic}[1]
	\Require $L, R \in \mathbb{R}^{n \times p}$
	\Comment{Observations}
	
	\Require $t^{(0)} \in \mathbb{R}^{m \times p}$
	\Comment{Initial support points}
	
	\For{$l \in \{ 1, \dots, B \}$}
		\State $\hat{g} \gets \arg\max_{\mathbf{w} \in \mathbb{R}^m_+ : \mathbf{1}'\mathbf{w}=1} \sum_{i=1}^n \log \sum_{k=1}^m w_k P(L_{i\cdot}, R_{i\cdot} \mid \theta_{i\cdot} = t^{(l-1)}_{k \cdot})$
		\State $\hat{\theta}^{(l)} \gets \hat{E}(\theta \mid L, R)$
		\State $\mu^{(l)}_1, \dots \mu^{(l)}_m \sim_{iid} \hat{g}$
		\State $t^{(l)}_k \mid \mu^{(l)}_k \sim N_p(\mu^{(l)}_k, \sigma^2 I_p)$
	\EndFor
	
	\State $\hat{\theta} \gets B^{-1} \sum_{l=1}^B \hat{\theta}^{(l)}$
\end{algorithmic}
\end{algorithm}

Algorithm \ref{alg} can be generalized to other empirical Bayes problems where $N(\mu, I_p)$ is replaced with another known likelihood. Additionally, a burn-in period of $K$ iterations can be included by simply ignoring the first $K$ iterations in the final estimation. In practice, we have found that $B=50$ provides a good balance between speed and estimation performance.

The empirical Bayes matrix estimation approach allows for many useful extensions. First, after the prior is estimated, it can be used to directly imputation and estimation of a new patient's biomarker values according to \eqref{eq:posteriormean}. Secondly, other posterior statistics, such as the mode and medoid, can be used to produce different estimates $\theta_{i\cdot}$ with properties such as sparsity. Additionally, statistics such as the posterior variance might provide a useful metric for providing weights based on confidence in down-stream learning tasks.

\section{Imputation Simulations}
\label{sec:sims}

We compared the performance of our method, EBM-Tobit, to other popular methods for censored biomarker measurement in simulations. Our simulation is based on the simulations used in previous missing not at random studies \cite{Wei_GSimp2018} and a bile acid dataset \cite{Lei_2017} previously used to study censored proteomics. The bile acid dataset contains the log-normal measurements of 34 bile acids for 198 patients; no missing values are present in the data. For each simulation, we generate $n = 1000$ patient biomarker measurements by first log-transforming the bile acid dataset so that it approximately follows a multivariate normal distribution. Next, we sample the true means, $\theta_{i \cdot}$, from a multivariate normal distribution whose mean and covariance match the empirical mean and covariance of $p = 25$ random bile acids in our dataset. Finally, for $\theta_{ij}$ falling below a pre-specified biomarker-specific quantile, $LOD_j$, an interval $[LB_j, LOD_j]$, where $LB_j = \min \theta_{\cdot j} - 6 \text{ sd}(\theta_{\cdot j})$ is observed. For $\theta_{ij}$ that are not censored, we observe one independent sample from $N(\theta_{ij}, \sigma_{ij}^2 = 1)$. We use a finite lower bound, $LB_j$, rather than $-\infty$, to avoid numerical issues in some of the methods; the log-normal interpretation of $LB_j$ is a very small, positive value. Note this simulation setting has at most one censoring interval per column, corresponding to the setting where each biomarker has a fixed lower detection limit.

The performance of our empirical Bayes matrix imputation method is compared to other popular imputation methods for missing not at random, left-censored data. The ``Tobit MLE'' method is maximum likelihood estimate defined in \eqref{eq:mle}; we note both that this method is a fill-in method in our simulation setting, and that this method simplifies to the $LOD/2$ fill-in method \cite{LeeKongWeissfeld2012} when the observed interval is $[0, LOD]$. ``QRILC'' \cite{QRILC} imputes the missing values using random draws from the estimated truncated normal distribution for each bile acid measured. The ``zCompositions'' method \cite{zCompositions} uses relative abondances to impute missing values. The default set-up of ``GSimp'' \cite{Wei_GSimp2018} imputes the missing values by repeatedly estimating the missing values using the fully observed data by repeatedly fitting an elastic-net model starting with the QRILC values. The ``trKNN'' method \cite{Shah_2017} is a nearest neighbors method applied by patient using the average of the nearest three patients' normalized, bile acid measurements to impute the missing values. Additionally, we include ``EB Oracle Support'' which denotes the nonparametric empirical Bayes $g$-modeling estimator, \eqref{eq:NPMLE}, using the optimal support points. This estimator cannot be computed in practice, because the optimal support points, $\theta_{i \cdot}$, are unknown, but it demonstrates that the methodology developed in Section \ref{sec:method} works well and that EBM-Tobit achieves performance reasonably close to optimal performance despite the difficulties with support point specification in this problem. 

\begin{figure}[ht]
	\centering
	\includegraphics{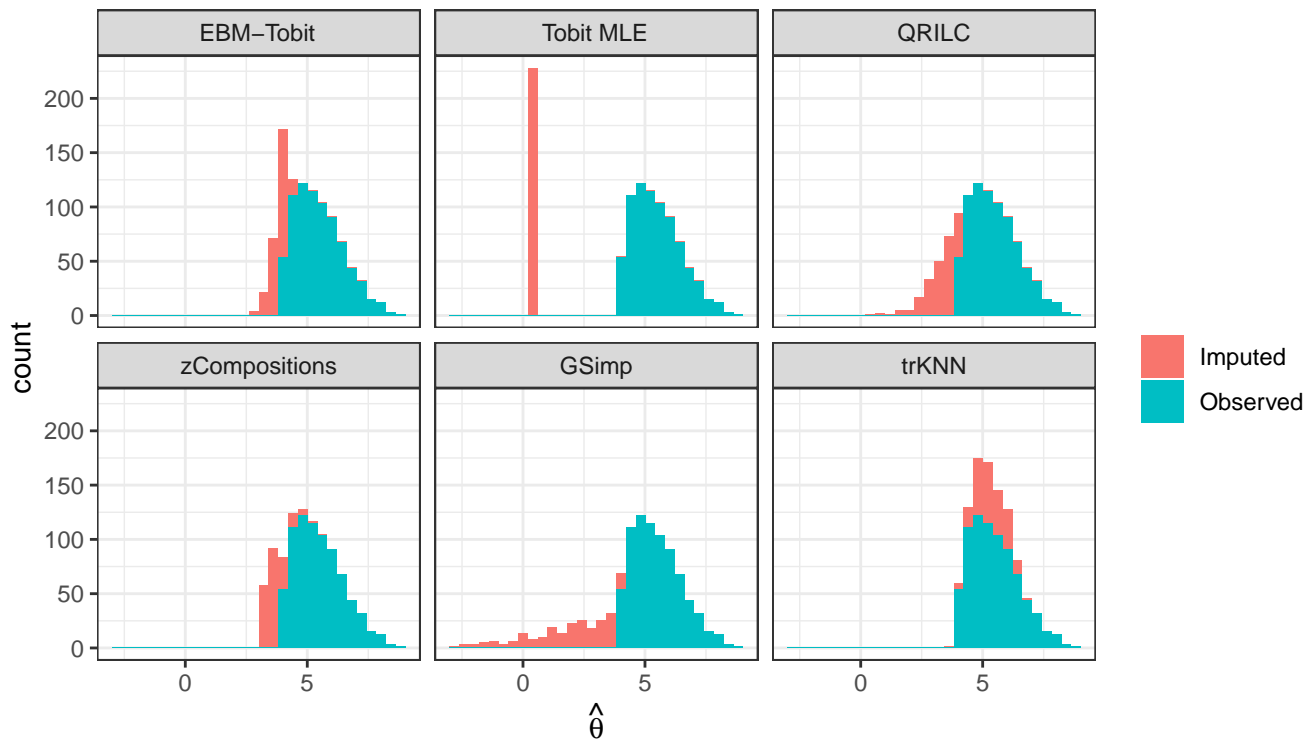}
	\caption{Each plot is the marginal histogram of a fixed, censored column. Three of ten columns are censored so that roughly 10\% of values below the detection limit. The ``EBM-Tobit'' histogram illustrates the our estimator from Algorithm \ref{alg} with $B=50$ iterations; ``Tobit MLE'' method is maximum likelihood estimate defined in \eqref{eq:mle}; ``QRILC'' is the typical QRILC method \cite{QRILC}; ``zCompositions'' is the log-normal zCompositions method \cite{zCompositions}; ``GSimp'' is the recommended version of GSimp \cite{Wei_GSimp2018}; and ``trKNN'' is the truncated K-nearest neighbors method \cite{Shah_2017}.}
	\label{fig:marginaldist}
\end{figure}

Figure \ref{fig:marginaldist} visualizes the marginal distributions produced by the imputation of the methods discussed above in one iteration of simulation where three of the ten columns have about 10\% of values below the detection limit. We know from the data generation process that the marginal distribution should be normal, so it is easy to see that QRILC does the best job capturing the marginal distribution, followed by our method, EBM-Tobit, and zCompositions. Our method appears to place more mass in the center of the histogram than QRILC while maintaining some lower tail, illustrating the shrinkage induced by the posterior mean. Furthermore, it is straightforward to see that the trKNN method is biased towards the observed data, GSimp is over-distributed, and the single value fill-in method, Tobit MLE, lacks variability that may make fitting down-stream methods difficult.

\begin{figure}[ht]
	\centering
	\includegraphics{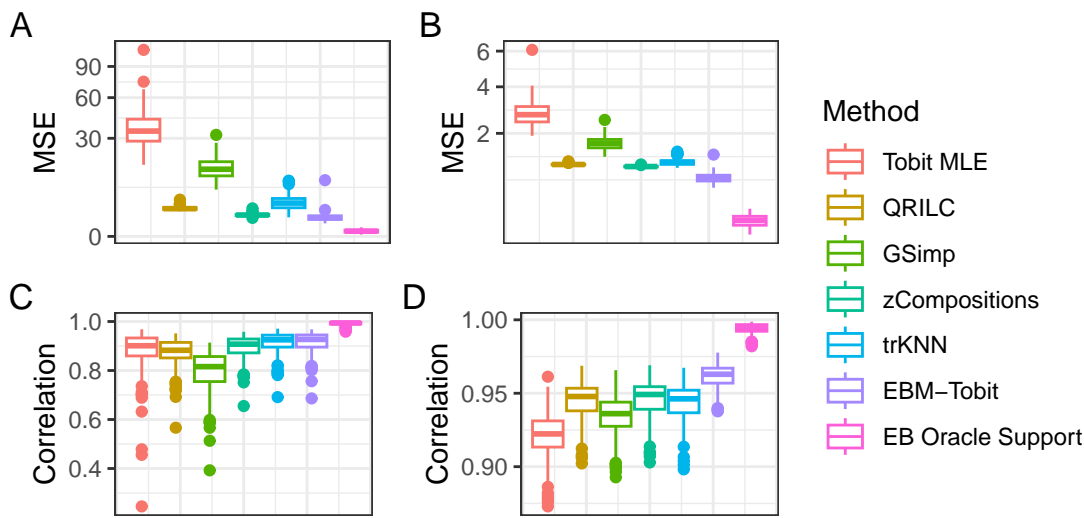}
	\caption{Plots comparing the performance of popular imputation methods for left-censored, missing not at random data to our empirical Bayes matrix estimation method. Plot A compares the mean squared error (on a square-root scale) computed only over $\theta_{ij}$ that have censored observations (imputation performance), while Plot B compares the the root mean squared error calculated over every $\theta_{ij}$ (estimation performance). Plots C and D show Spearman's correlation over the same $\theta_{ij}$ as Plots A and B. The methods are as follows: ``QRILC'' is the typical QRILC method \cite{QRILC}; ``GSimp'' is the recommended version of GSimp \cite{Wei_GSimp2018}; ``zCompositions'' is the log-normal zCompositions method \cite{zCompositions}; ``trKNN'' is truncated K-nearest neighbors method \cite{Shah_2017}; ``Tobit MLE'' method is maximum likelihood estimate defined in \eqref{eq:mle}; ``EBM-Tobit'' denotes our estimator from Algorithm \ref{alg} using $B=50$ iterations; and ``EB Oracle Support'' is the nonparametric empirical Bayes $g$-modeling estimator using the optimal support points.}	
	\label{fig:mse_comarison}
\end{figure}

We empirically compare the performance of these imputation methods across 200 rounds of simulations. The dimension of the problem is fixed at $n=1000$ samples and $p=25$ bile acids and eight of the bile acids have approximately 10\% left-censored measurements. Simulations covering different number of censored columns and different levels of censoring are left to Appendix \ref{app:sims}. Because we are interested in both imputation performance and the ability to estimate the whole matrix, we measure root mean squared error and Spearman's correlation over just the censored values as well as over every value. The metrics are computed with respect to the simulated, true means. Results are visualized in Figure \ref{fig:mse_comarison}. 

These simulation results demonstrate that our empirical Bayes matrix estimation method, EBM-Tobit, frequently matches the best imputation performance of popular methods for left-censored, missing not at random data. Moreover, EBM-Tobit greatly outperforms the other methods for whole matrix estimation. We note that zCompositions, which performs as well as EBM-Tobit in Figure \ref{fig:mse_comarison} Plots A and C, is only applicable to left-censored problems. We additionally note that the oracle empirical Bayes method vastly outperforms popular imputation methods in all simulations, offering strong justification our empirical Bayes approach.

\section{Discussion}
\label{sec:disc}

One of the key advantages of empirical Bayes methods is their ability to induce shrinkage in the estimation problem. By leveraging a data-dependent prior distribution, empirical Bayes methods borrow information across multiple observations and produce more stable and reliable parameter estimates. Figure \ref{fig:mse_comarison} illustrates that our empirical Bayes estimates are consistently close to the true means and captures variability that is likely to help improve down-stream analysis with tools designed for continuous inputs. We note that because EBM-Tobit is designed to estimate all of the true means, not just the censored ones, it is the only method to have an mean squared error less than one when estimating all of the means.

Our methodology has been focused on the class of all priors on $\mathbb{R}^p$, allowing for arbitrary dependence between biomarker values. This dependence between biomarker values is different than modeling correlated measurement errors and is closer to learning the true physical model for the biological processes. However, in many applications there may be additional domain knowledge that can be incorporated as restrictions on the space of priors. For example, if various sets of biomarkers are known to be unrelated, a corresponding independence structure can be imposed on the class of priors. This allows the estimation problem to be bifurcated, both decreasing the difficulty of each sub-problem and allowing for parallelization of model fitting. Additionally, the support of the prior can be restricted to incorporate knowledge of the biomarker's support, such as non-negativity. By restricting the space of priors, we produce more efficient estimators.

Empirical Bayes models are often discussed in the context of shrinkage estimators. In this case, it is pertinent to ask ``where are we shrinkage to?'' Since our application mainly concerns imputing left-censored means a reasonable question is: should we shrink towards the global mean given that we know the observation was on the low end? This Efron's relevance problem \cite{Efron2019}. It is not necessary that $\theta_{ij}$ lies in $[L_{ij}, R_{ij}]$; however, in the case of detection limits, the fact that a measurement is censored still somewhat informative. This suggests it may be good to include the information that the observation was censored in the estimation procedure. One simple solution is to define a known covariate to indicate whether the observation was censored. Including this binary covariate into the empirical Bayes model results in estimating two separate priors and corresponding posteriors. Because we are partitioning our data in this approach, the estimation of each prior becomes less efficient; for this reason, it may be better to bet on the flexibility of the nonparametric prior we are already using to adapt to these sub-populations especially when the sub-populations are small or our domain expertise is limited.

\bibliography{paper}

\appendix

\section{Additional Simulations}
\label{app:sims}

In this we extend the simulations from Section \ref{sec:sims} to different types and degrees of missingness. We find that our estimator computed with Algorithm \ref{alg}, EBM-Tobit, generally performs similarly to other methods at imputation while out-performing other methods at estimating the whole matrix. Two metrics (mean squared error and Spearman correlation) and two problems (imputation and estimation) are considered across the following four tables. The values reported in the tables are the average over 200 simulations, as described in Section \ref{sec:sims}. Because the simple fill-in methods only impute a single value, based on the common lower detection limit, there may be missing correlation values where the correlation cannot be computed.

\begin{table}

\caption{\label{tab:mse-cen}Imputation performance measured by mean squared error over censored means only; averaged over 200 simulations.}
\centering
\begin{tabular}[t]{l|r|r|r|r|r|r|r|r|r}
\hline
\multicolumn{1}{c|}{Percent Missing Columns} & \multicolumn{3}{c|}{10\%} & \multicolumn{3}{c|}{30\%} & \multicolumn{3}{c}{50\%} \\
\cline{1-1} \cline{2-4} \cline{5-7} \cline{8-10}
\multicolumn{1}{c|}{Lower Detection Limit Quantile} & \multicolumn{1}{c|}{0.1} & \multicolumn{1}{c|}{0.3} & \multicolumn{1}{c|}{0.5} & \multicolumn{1}{c|}{0.1} & \multicolumn{1}{c|}{0.3} & \multicolumn{1}{c|}{0.5} & \multicolumn{1}{c|}{0.1} & \multicolumn{1}{c|}{0.3} & \multicolumn{1}{c}{0.5} \\
\cline{1-1} \cline{2-2} \cline{3-3} \cline{4-4} \cline{5-5} \cline{6-6} \cline{7-7} \cline{8-8} \cline{9-9} \cline{10-10}
QRILC & 2.591 & 2.712 & 3.353 & 2.457 & 2.727 & 3.263 & 2.452 & 2.730 & 3.224\\
\hline
GSimp & 15.570 & 5.658 & 2.008 & 14.757 & 6.204 & 1.979 & 15.382 & 6.070 & 2.028\\
\hline
zCompositions & 1.500 & 1.446 & 1.982 & 1.443 & 1.424 & 1.984 & 1.435 & 1.428 & 2.006\\
\hline
trKNN & 3.866 & 5.016 & 6.877 & 3.751 & 5.427 & 6.775 & 4.099 & 5.605 & 6.939\\
\hline
Half-Min & 6.174 & 5.713 & 6.072 & 5.519 & 5.775 & 5.973 & 5.357 & 5.840 & 5.922\\
\hline
EBM-Tobit & 1.221 & 1.485 & 2.301 & 1.176 & 1.386 & 2.679 & 1.137 & 1.421 & 2.556\\
\hline
Generalized Exemplar Support & 39.938 & 43.085 & 47.006 & 36.539 & 42.277 & 44.937 & 36.014 & 42.998 & 44.223\\
\hline
Oracle Support Points & 0.083 & 0.130 & 0.187 & 0.094 & 0.177 & 0.244 & 0.096 & 0.191 & 0.327\\
\hline
Vectorized Oracle & 6.756 & 6.452 & 5.810 & 6.981 & 6.526 & 6.527 & 7.155 & 6.800 & 6.865\\
\hline
\end{tabular}
\end{table}

\begin{table}

\caption{\label{tab:mse-all}Estimation performance measured by mean squared error over all means; averaged over 200 simulations.}
\centering
\begin{tabular}[t]{l|r|r|r|r|r|r|r|r|r}
\hline
\multicolumn{1}{c|}{Percent Missing Columns} & \multicolumn{3}{c|}{10\%} & \multicolumn{3}{c|}{30\%} & \multicolumn{3}{c}{50\%} \\
\cline{1-1} \cline{2-4} \cline{5-7} \cline{8-10}
\multicolumn{1}{c|}{Lower Detection Limit Quantile} & \multicolumn{1}{c|}{0.1} & \multicolumn{1}{c|}{0.3} & \multicolumn{1}{c|}{0.5} & \multicolumn{1}{c|}{0.1} & \multicolumn{1}{c|}{0.3} & \multicolumn{1}{c|}{0.5} & \multicolumn{1}{c|}{0.1} & \multicolumn{1}{c|}{0.3} & \multicolumn{1}{c}{0.5} \\
\cline{1-1} \cline{2-2} \cline{3-3} \cline{4-4} \cline{5-5} \cline{6-6} \cline{7-7} \cline{8-8} \cline{9-9} \cline{10-10}
QRILC & 1.012 & 1.042 & 1.095 & 1.045 & 1.171 & 1.362 & 1.068 & 1.256 & 1.534\\
\hline
GSimp & 1.171 & 1.121 & 1.041 & 1.679 & 1.549 & 1.156 & 2.063 & 1.802 & 1.247\\
\hline
zCompositions & 0.998 & 1.008 & 1.040 & 0.993 & 1.028 & 1.157 & 0.990 & 1.043 & 1.243\\
\hline
trKNN & 1.024 & 1.102 & 1.236 & 1.109 & 1.463 & 1.923 & 1.191 & 1.723 & 2.425\\
\hline
Half-Min & 1.051 & 1.121 & 1.204 & 1.199 & 1.501 & 1.795 & 1.286 & 1.761 & 2.181\\
\hline
EBM-Tobit & 0.729 & 0.753 & 0.784 & 0.724 & 0.766 & 1.021 & 0.726 & 0.810 & 1.130\\
\hline
Generalized Exemplar Support & 1.454 & 2.119 & 2.845 & 2.783 & 5.471 & 8.029 & 3.630 & 7.814 & 11.371\\
\hline
Oracle Support Points & 0.091 & 0.091 & 0.102 & 0.104 & 0.123 & 0.147 & 0.107 & 0.134 & 0.198\\
\hline
Vectorized Oracle & 0.960 & 1.036 & 1.082 & 1.183 & 1.489 & 1.782 & 1.341 & 1.831 & 2.310\\
\hline
\end{tabular}
\end{table}

\begin{table}

\caption{\label{tab:cor-cen}Imputation performance measured by Spearman correlation over censored means only; averaged over 200 simulations.}
\centering
\begin{tabular}[t]{l|r|r|r|r|r|r|r|r|r}
\hline
\multicolumn{1}{c|}{Percent Missing Columns} & \multicolumn{3}{c|}{10\%} & \multicolumn{3}{c|}{30\%} & \multicolumn{3}{c}{50\%} \\
\cline{1-1} \cline{2-4} \cline{5-7} \cline{8-10}
\multicolumn{1}{c|}{Lower Detection Limit Quantile} & \multicolumn{1}{c|}{0.1} & \multicolumn{1}{c|}{0.3} & \multicolumn{1}{c|}{0.5} & \multicolumn{1}{c|}{0.1} & \multicolumn{1}{c|}{0.3} & \multicolumn{1}{c|}{0.5} & \multicolumn{1}{c|}{0.1} & \multicolumn{1}{c|}{0.3} & \multicolumn{1}{c}{0.5} \\
\cline{1-1} \cline{2-2} \cline{3-3} \cline{4-4} \cline{5-5} \cline{6-6} \cline{7-7} \cline{8-8} \cline{9-9} \cline{10-10}
QRILC & 0.601 & 0.551 & 0.525 & 0.874 & 0.862 & 0.824 & 0.898 & 0.883 & 0.859\\
\hline
GSimp & 0.564 & 0.622 & 0.669 & 0.797 & 0.861 & 0.881 & 0.817 & 0.876 & 0.903\\
\hline
zCompositions & 0.618 & 0.601 & 0.581 & 0.895 & 0.897 & 0.870 & 0.921 & 0.911 & 0.896\\
\hline
trKNN & 0.796 & 0.771 & 0.757 & 0.913 & 0.910 & 0.892 & 0.925 & 0.915 & 0.905\\
\hline
Half-Min & 0.726 & 0.681 & 0.668 & 0.907 & 0.907 & 0.879 & 0.927 & 0.921 & 0.906\\
\hline
EBM-Tobit & 0.717 & 0.704 & 0.699 & 0.916 & 0.919 & 0.898 & 0.934 & 0.931 & 0.921\\
\hline
Generalized Exemplar Support & 0.665 & 0.627 & 0.615 & 0.883 & 0.877 & 0.844 & 0.904 & 0.886 & 0.867\\
\hline
Oracle Support Points & 0.981 & 0.973 & 0.970 & 0.992 & 0.989 & 0.985 & 0.993 & 0.990 & 0.985\\
\hline
Vectorized Oracle & 0.727 & 0.686 & 0.673 & 0.906 & 0.906 & 0.882 & 0.926 & 0.920 & 0.907\\
\hline
\end{tabular}
\end{table}

\begin{table}

\caption{\label{tab:cor-all}Estimation performance measured by Spearman correlation over all means; averaged over 200 simulations.}
\centering
\begin{tabular}[t]{l|r|r|r|r|r|r|r|r|r}
\hline
\multicolumn{1}{c|}{Percent Missing Columns} & \multicolumn{3}{c|}{10\%} & \multicolumn{3}{c|}{30\%} & \multicolumn{3}{c}{50\%} \\
\cline{1-1} \cline{2-4} \cline{5-7} \cline{8-10}
\multicolumn{1}{c|}{Lower Detection Limit Quantile} & \multicolumn{1}{c|}{0.1} & \multicolumn{1}{c|}{0.3} & \multicolumn{1}{c|}{0.5} & \multicolumn{1}{c|}{0.1} & \multicolumn{1}{c|}{0.3} & \multicolumn{1}{c|}{0.5} & \multicolumn{1}{c|}{0.1} & \multicolumn{1}{c|}{0.3} & \multicolumn{1}{c}{0.5} \\
\cline{1-1} \cline{2-2} \cline{3-3} \cline{4-4} \cline{5-5} \cline{6-6} \cline{7-7} \cline{8-8} \cline{9-9} \cline{10-10}
QRILC & 0.946 & 0.944 & 0.943 & 0.945 & 0.941 & 0.936 & 0.943 & 0.939 & 0.929\\
\hline
GSimp & 0.943 & 0.943 & 0.944 & 0.935 & 0.936 & 0.942 & 0.927 & 0.931 & 0.938\\
\hline
zCompositions & 0.946 & 0.946 & 0.944 & 0.946 & 0.944 & 0.943 & 0.945 & 0.944 & 0.939\\
\hline
trKNN & 0.946 & 0.942 & 0.936 & 0.944 & 0.933 & 0.917 & 0.942 & 0.927 & 0.905\\
\hline
Half-Min & 0.945 & 0.942 & 0.939 & 0.942 & 0.934 & 0.925 & 0.940 & 0.929 & 0.917\\
\hline
EBM-Tobit & 0.961 & 0.960 & 0.959 & 0.961 & 0.960 & 0.956 & 0.961 & 0.959 & 0.953\\
\hline
Generalized Exemplar Support & 0.940 & 0.929 & 0.924 & 0.921 & 0.889 & 0.863 & 0.907 & 0.865 & 0.829\\
\hline
Oracle Support Points & 0.995 & 0.995 & 0.994 & 0.994 & 0.993 & 0.992 & 0.994 & 0.993 & 0.990\\
\hline
Vectorized Oracle & 0.942 & 0.936 & 0.934 & 0.930 & 0.912 & 0.898 & 0.921 & 0.895 & 0.873\\
\hline
\end{tabular}
\end{table}

In addition to comparing our methods to existing methods, we include a few alternative empirical Bayes methods to help measure the performance of our estimator. Specifically, we include ``Generalized Exemplar Support'' which uses \eqref{eq:mle} as the support points for the prior; the results suggest that EBM-Tobit generally does better. We also include oracle empirical Bayes procedures that use cannot be calculated in practice, but surve as a baseline for our methods. First we include ``Oracle Support Points'' which uses the optimal support points: $\theta_{i \cdot}$, this method does very well. We also include ``Vectorized Oracle'' which treats the problem as a vector denoising problem rather than a matrix denoising problem; after vectorizing we have $\theta_{ij} \sim g$ for some $g$ on $\mathbb{R}$, again, we use the optimal support points. The large performance gap between ``Oracle Support Points'' and ``Vectorized Oracle'' indicates that the matrix structure is very useful for this estimation problem.

\section{Benefits of Multivariate Priors}
\label{app:circle-demo}

\begin{figure}[ht]
	\centering
	\includegraphics{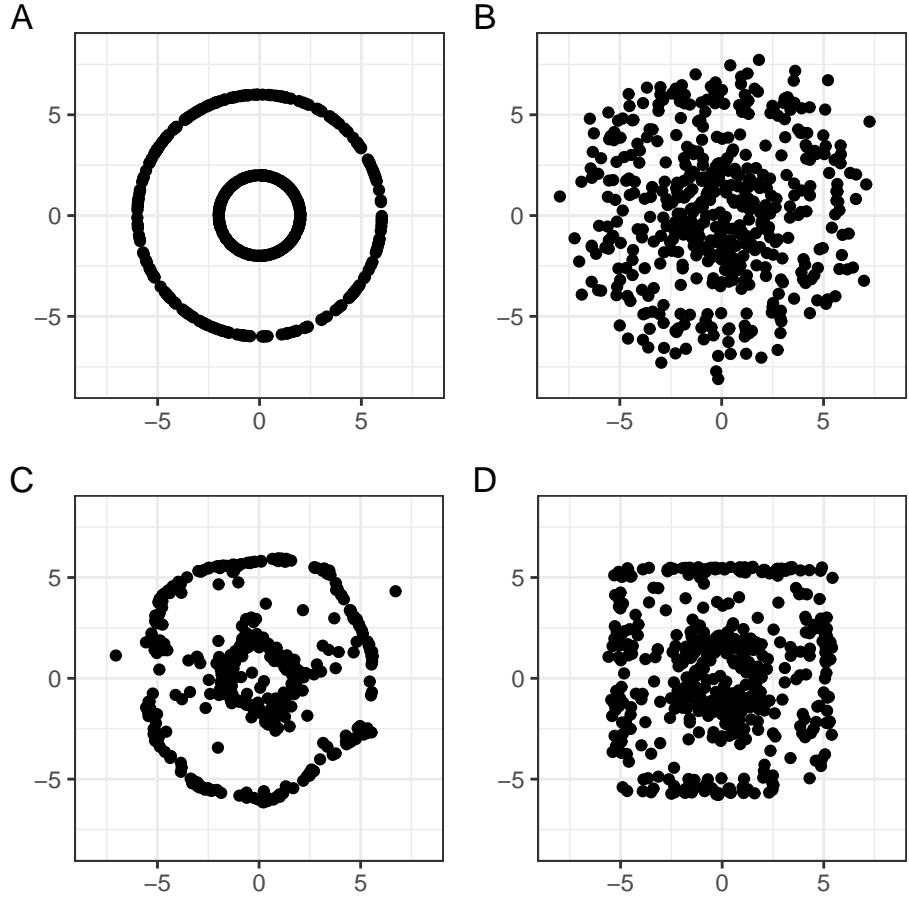}
	\caption{An illustration of the ability of joint prior to encode complex relationships between variables. Plot A shows $n = 500$ samples from the oracle prior: uniform over two concentric circles with radii 2 and 6. Plot B shows the observations drawn independently as $x_i \sim N_2(\theta_{i 1}, 1)$ and $y_i \sim N_2(\theta_{i 2}, 1)$, for $i = 1, \dots, n$. Plots C and D show the estimated posterior means resulting from the exemplar method using a joint prior on $\mathbb{R}^2$ and two independent priors on $\mathbb{R}$, respectively. We note that Plot C looks much more circular, like the true means in Plot A, than Plot D.}
	\label{fig:circle}
\end{figure}

Figure \ref{fig:circle} demonstrates the ability of an arbitrary prior to encode complicated interactions. This figure illustrates that a prior with complicated joint relationships cannot be properly recovered when an mean field structure (indepence assumption) improperly imposed. This is seen readily in the lack of corners in Plot C compared to Plot D. This demonstration is based on Figure 1 of Saha and Guntuboyina (2020) \cite{SahaGuntuboyina2020}.

\end{document}